\documentstyle[prb,aps,galley,floats,psfig]{revtex}
\newlength{\figwidth}
\newlength{\refspace}

\renewcommand{\thesection}{\Roman{section}}
\renewcommand{\thesubsection}{\Alph{subsection}}

\begin{document}
\newcommand{\be}{\begin{equation}}
\newcommand{\ee}{\end{equation}}
\def\th{\theta}
\def\tc{\theta^*}
\def\Tc{\theta^{**}}
\def\Lm{\Lambda}
\def\lm{\lambda}
\def\Gm{\Gamma}
\def\c2{{\rm cn^2}}
\def\s2{{\rm sd^2}}
\def\Lamdar{\Lambda_{\rm R}}
\def\half{\frac{1}{2}}
\def\reva{\frac{1}{4}}
\def\ints{\int_0^{2\pi}}
\def\inth{\int_0^{\pi}}
\def\ra{\rightarrow}
\author{Jorge Berger$^o$ and  Jacob Rubinstein$^*$ \\
$^o$Department of Physics, Technion, 32000 Haifa, Israel\\
$^*$Department of Mathematics, Technion, 32000 Haifa, Israel}
\title{Design for the Detection of the Singly-Connected 
Superconducting State}
\maketitle
{
\begin{abstract}
We study the Little-Parks effect for mesoscopic loops with very nonuniform
thickness. The results follow the trend of the phase diagram obtained for
almost uniform thickness. In particular, the singly-connected state is stable
on a line segment delimited by two critical points. Most of this study
considers loops with piecewise constant thickness; in this case the 
Euler-Lagrange equation can be integrated analytically. Under appropriate 
conditions, the temperature range where the singly-connected state 
is stable is proportional to the square of the ratio between the maximal and
the minimal thicknesses. Our results may serve as a guide for
planning experiments.
\end{abstract}     } {
}
\narrowtext

\section{INTRODUCTION}
\begin{figure}[tbh]
\centerline{\psfig{figure=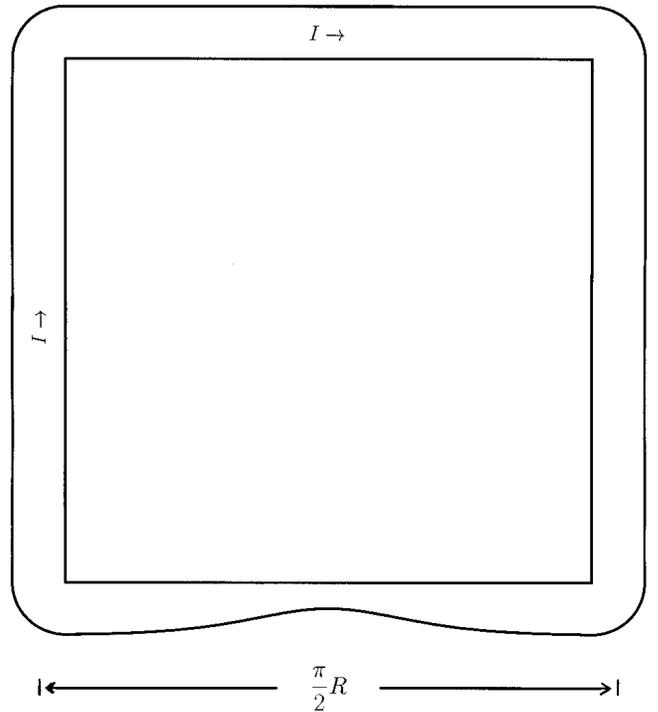,width=\figwidth}}
\caption{ Thin loop of superconducting material with perimeter $2\pi R$
and
nonuniform cross section $D$. The magnetic flux enclosed by the loop is 
$\Phi$ and the supercurrent around it is $I$.}
\label{fig1}
\end{figure}

We deal with a loop of superconducting material as shown in
Fig.~\ref{fig1}. If the
entire loop is superconducting and if it encloses a non-integer number of
magnetic flux quanta, then single-valuedness of the order parameter requires
the presence of some supercurrent $I$. This supercurrent involves an energy
price and, due to it, the transition to superconductivity occurs at lower 
temperatures when the enclosed flux is non-integer. This effect was observed
by Little and Parks \cite{lipa} and an explanation in terms of the 
Ginzburg-Landau theory was provided by Tinkham.\cite{tin}

We have recently predicted that, if the thickness of the loop is not exactly
uniform, then there exist situations for which superconductivity is broken
at a layer, so that the superconducting part is actually singly-connected
\cite{beru} and no supercurrent flows.
When this happens, we say that the sample is in the ``singly-connected state"
(SC). This is an interesting possibility, since it would allow for  a new 
dimensionality of the regions where the order parameter may vanish.
In the case of vortices, the order parameter vanishes along lines and there
are claims \cite{friend} that it cannot vanish on surfaces. 
On the other hand, it has been suggested that even for uniform thickness the 
SC state will appear as an intermediate station in hysteretic
 paths.\cite{gust} A systematic
analytic study for the stability domain of the SC state
in families of loops with thicknesses that deviate slightly from uniformity
was carried on in Ref.~\onlinecite{SIAM}. Mathematical justification for
some of the 
assumptions in our model is given in Ref.\ \onlinecite{Mich}. Another
situation in 
which the order parameter seems to vanish on a layer was considered in
Refs.\ \onlinecite{fink} -- \onlinecite{moch1}.

Since Ref.\ \onlinecite{SIAM} may be too mathematically oriented, we
shall review
here its central findings. We assume that the thickness of the loop (radial
dimension) is
much smaller than the perimeter. In this case, for the range of temperatures
at which transitions occur, the order parameter is uniform on cross
sections of the loop. (A ``cross section" is a surface which is everywhere 
normal to the supercurrent.) 
In addition, the free energy is insensitive to the magnetic field created by 
$I$ and to the linear shape of the loop. We may therefore regard
the loop as a circular ring with radius $R$ equal to the perimeter divided by
$2\pi$. We define by $\th$ the angle which is obtained by deforming
the loop to a circle, i.e. $\th$ is the arc length divided by $R$; 
the ``thickness" $D(\th)$ is defined
as the area of the cross section at $\th$. The origin $\th=0$ is defined by 
requiring that
\be
\ints D(\th)\sin\th d\th=0
\label{zero}
\ee
 and
\be
\beta=2\ints D(\th) \cos\th d\th\left/\ints D(\th) d\th <0 \right. \ .
\label{beta}
\ee

 The controllable physical 
coordinates are the temperature $T$ and the magnetic flux $\Phi$. They enter
our equations through
\be
\lm=\frac{R^2}{\xi^2}=\frac{T_c-T}{T_c-T_R}
\label{lm}
\ee
and
\be
k=n-\Phi/\Phi_0,
\label{k}
\ee
where $\xi$ is the coherence length, $T_c$ is the critical temperature in
the absence of magnetic field, $T_R$ is the temperature at which $\xi=R$,
$\Phi_0$ is the quantum of magnetic flux and $n$ is some integer. 

Let $\psi(\th)$ be the order parameter and $\psi_0$ the order parameter
that would be obtained in the absence of 
magnetic field. We define $y(\th)=|\psi(\th)/\psi_0|$. 
The contribution of superconductivity to the free energy is proportional to 
\be
 \int_0^{2\pi} (-\lm y^2+\half \lm y^4 +(y^{'})^2 )
D d \th +(2\pi k)^2\Lm^{-1} \ ,
\label{gl3}
\ee
where the prime denotes differentiation with respect to $\th$ and
\be
\Lm= \ints \frac{d\th}{Dy^2}
\label{Lm}
\ee
is a nonlocal term. (Some symbols, such as $\Lm$ and $\lm_i$ are defined
here not exactly as in Ref.\ \onlinecite{SIAM}.) 
The term $k/\Lm$ (cf. (\ref{gl3})) is proportional to the 
supercurrent $I$. Since (\ref{gl3}) increases monotonically with $k^2$,
the minimum of the free energy  will always occur for
$k$ in the range $[-\half,\half]$.
Note that only the absolute value of $\psi(\th)$ enters the expression for
the the free energy; $\arg[\psi(\th)]$ has been already worked out.

A necessary condition for a minimum of the free energy is the 
Euler-Lagrange (EL) equation. 
For the SC state $y(0)=0$ and the EL equation is
\be
(Dy^{'})^{'}+\lm D(y-y^3)=0 \ .
\label{ely}
\ee
When $y(\th)$ is positive everywhere, we say that the sample is in the
``doubly-connected state" (DC). In this case a better description is given in 
terms of $w=y^2$ and the EL equation reads
\begin{eqnarray}
Dww^{''}+D^{'}ww^{'}-\frac{D}{2}(w^{'})^2+2\lm Dw^2(1-w) \nonumber \\
-\frac{2}{D}\left(\frac{2\pi k}{\Lm}\right)^2 =0.
\label{elw}
\end{eqnarray}

\begin{figure}[tbh]
\centerline{\psfig{figure=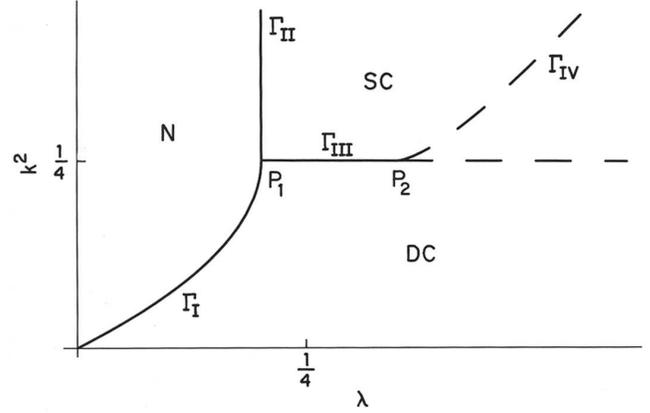,width=\figwidth}}
\caption{Phase diagram. N: $y \equiv 0$; DC: $y>0$ everywhere; SC: the order
parameter vanishes at some place in the loop and becomes singly connected.
P$_1$ and P$_2$ are critical points. $\lm$ decreases with temperature and
$k^2$ increases with the deviation from an integer number of flux quanta.}
\label{fig2}
\end{figure}

The phase diagram in the temperature - magnetic field plane is shown in 
Fig.~\ref{fig2}. N, DC and SC denote the three possible states and
$\Gm_{\rm I}$,
$\Gm_{\rm II}$ and $\Gm_{\rm III}$ are critical lines at which second order 
phase transitions occur. When $\Gm_{\rm III}$ is approched from below, 
$y(0)$ decreases until it finally vanishes. Since $\Gm_{\rm III}$ is 
located along $|k|=\half$, and the most stable state is never at $|k|>\half$, 
the SC stability domain is restricted to a line segment. The 
ends of this segment are the critical points P$_1$ and P$_2$. When the 
magnetic field is varied and the line $|k|=\half$ is crossed for small
$\lm$ (between P$_1$ and P$_2$), the current $I$ vanishes and changes sign 
continuously. However, for large $\lm$ (beyond P$_2$), the current $I$ 
changes sign discontinuously.

Let us denote the positions of P$_{1,2}$ by $(\lm=\lm_{1,2},|k|=\half)$.
$\lm_1$ is given by the smallest value $\lm$ for which the 
linearized Eq.~(\ref{ely}) (without $y^3$) has nontrivial solutions. To 
locate $\lm_2$ we define
\be
f(\th)=D(\th)w(\th), \;\; \tilde{f}(\th)=f(0)+\half f''(0)\th^2,
\label{f}
\ee
\be
\Lm_{\rm R}=-2\int_{\pi}^{\infty}\frac{d\th}{\tilde{f}}+
2\int_0^{\pi}\frac{\tilde{f}-f}{f\tilde{f}}d\th.
\label{Lamdar}
\ee
Then, along $\Gm_{\rm III}$, $\Lm_{\rm R}$ is negative (resp. zero, positive) 
when the value of
$\lm$ is less (resp. equal, larger) than $\lm_2$. If $D(\th)$ is 
close to uniform, then, to first order in the deviation from uniformity, 
\be
\lm_1=\reva(1+\beta); \; \; \lm_2=\reva(1-2\beta) \ .
\label{define}
\ee

The analysis of Ref.\ \onlinecite{SIAM} has an obvious shortcoming
for experimental
purposes: one would like the SC state to exist in a temperature
range which is not too small, i.e. P$_1$ and P$_2$ are not too close. This
occurs when $|\beta|$ is not too small. But most of the results reviewed 
above were obtained under the assumption that $D(\th)$ is almost uniform,
which implies $|\beta|<<1$. The purpose of this article is to study forms
of $D(\th)$ which are strongly nonuniform, to check whether the scenario
depicted in Fig.~\ref{fig2} is still valid and whether only the first
harmonic of
$D(\th)$ affects the phase diagram [since $D(\th)$ enters Eq.~(\ref{define})
only through $\beta$]. 

In most of this article (Secs.~II and III) we study the case in which 
$D(\th)$ is a
piecewise constant function. In this case Eq.~(\ref{elw}) can be solved
analytically. Sec.~III considers the experimentally interesting limit
of a weak link. We shall see that in this case P$_2$ can move to temperatures
where superconductivity is already well developed, provided that the ratio
between the length and the width of the weak link is carefully tuned. In
Sec.~IV we consider deviations from the piecewise constant shape and Sec.~V
provides suggestions for experiments which, besides the appropriate form for
$D(\th)$, are quite standard.

\section{PIECEWISE CONSTANT THICKNESS}

A piecewise constant thickness will be described by $D(\th)$ of the form
\be
D=\left\{ \begin{array}{ll}
d & \;\;\th < \tc \\
1 & \;\;\tc < \th < \pi
\end{array} \right.
\label{piece}
\ee
with $0 < d < 1$ and $0<\tc<\pi$ constants, and $D(\th)$ symmetric 
about $\th=0$ and about $\th=\pi$. We are free to set the maximum of $D(\th)$
as 1, since the EL equation is invariant under multiplication of $D(\th)$ by
a constant; $d$ is the ratio between the minimum and the maximum thicknesses.

In this case we have found a solution of Eq.~(\ref{elw}) with the 
appropriate symmetry. For $\th<\tc$
we require the order parameter to have a minimum at $\th=0$. It has the form
\be
w(\th)=A_1-\frac{2\nu_1^2}{\lm}m_1 \c2(\nu_1 \th,m_1).
\label{l*}
\ee
For $\th>\tc$ we require a maximum at $\th=\pi$ and the solution is
\be
w(\th)=A_2-\frac{2\nu_2^2}{\lm}m_2(1-m_2) \s2(\nu_2 \th,m_2).
\label{m*}
\ee
In these expressions cn and sd are Jacobian elliptic functions, \cite{abi}
$\nu_i$ and $m_i$ are constants, and 
\be
A_i=\frac{2}{3}\left(1+(2m_i-1)\frac{\nu_i^2}{\lm}\right).
\label{Ai}
\ee
With these forms, the constant term in Eq.~(\ref{elw}) becomes
\be
\left(\frac{2\pi k}{D\Lm}\right)^2 =
\frac{A_i}{8\lm}\left(\lm^2(A_i-2)^2-4\nu_i^4\right) \ .
\label{elcons}
\ee

\begin{figure}[tbh]
\centerline{\psfig{figure=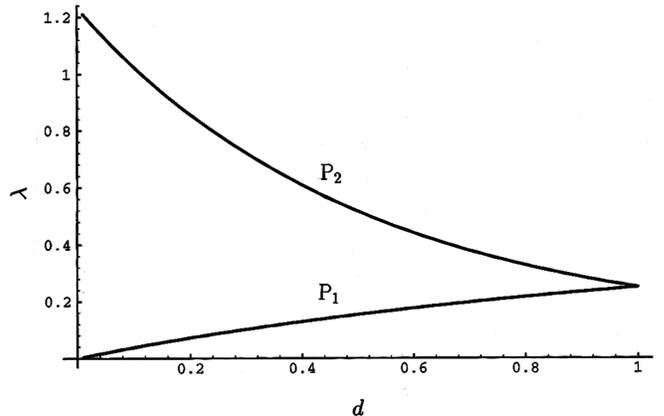,width=\figwidth}}
\caption{Position of the critical points as functions of the ratio 
$d=D(0)/D(\pi)$ in (\protect{\ref{piece}}). Here $\tc=\pi/2$.}
\label{fig3}
\end{figure}

\subsection{Mathematical Details}

Eqs.~(\ref{l*})--(\ref{elcons}) reduce the integro-differential equation 
(\ref{elw}) to the problem of determining the four constants $\nu_i$ and
$m_i$. We are thus left with four algebraic equations:
Eq.~(\ref{elcons}) for $i=1,2$ and continuity of $w$ and $Dw'$ at
 $\th=\tc$.\cite{SIAM} These equations are solved by Newton iterations. 
The integration in (\ref{Lm}) is performed numerically.

When $w(0)=0$, $\Lm$ diverges and Eqs.~(\ref{elcons}) simplify to 
$\lm(2-A_i)=2\nu_i^2$.
Still for $w(0)=0$, the numerator and the denominator of the integrand in
(\ref{Lamdar}) vanish for $\th\ra 0$.
In this region, we expanded the integrand (analytically) into a power series.

It is sometimes useful to regard $w(0)$ as an independent variable (and $k$
as dependent). For $0<w(0)<<1$, the integration in (\ref{Lm}) cannot be
performed numerically near $\th=0$. 
In this region, we substitute the integrand by a Pad\'e approximant in powers 
of $\th^2$, with the denominator linear in $\th^2$.

As $\Gm_{\rm I}$ is approached, $m_{1,2}\ra 0$, and we recover the 
situation in subsection 4.3 of Ref.\ \onlinecite{SIAM}. As $\lm$ is
increased away
from $\Gm_{\rm I}$, $m_1$ and $m_2$ increase, but $m_1$ increases faster than
$m_2$. Not far from $\Gm_{\rm I}$, $m_1=1$. If $\lm$ is increased further, 
$m_1$ and $\nu_1$ become complex numbers, with $|m_1|=1$ and 
$\arg(\nu_1)=-\reva\arg(m_1)$. Both regimes ($m_1$ and $\nu_1$ real or 
complex) can be unified by writing $m_1=e^{-4t}$ and $\nu_1=re^t$:
Eqs.~(\ref{l*}) and (\ref{Ai}) together with relationships 16.9.1 and
16.11.3 of Ref.\ \onlinecite{abi} show that $w$ is an even function
of $t$. 
It follows that $w(\th)$ depends on $t$ only through $t^2$; more precisely,
$w(\th)$ is an analytic function of $t^2$ which is real when $t^2$ is real,
and nothing special happens when $t^2$ changes sign. This means that
$w$ is still real when $m_1$ and $\nu_1$ become complex, and there is no
singularity at $m_1=1$. Our computer programs used different
parametrizations for the cases $0<m_1<1$ and $|m_1|=1$. Since we are mostly
interested in the region where $\lm$ is large, in some cases we did not 
perform calculations in the region where $0<m_1<1$.

\subsection{Dependence of $\lm_{1,2}$ on $d$}

By a straightforward generalization of subsection 4.3 in 
Ref.\ \onlinecite{SIAM}, we obtain
\be
d=\tan(\sqrt{\lm_1}(\pi-\tc))\tan(\sqrt{\lm_1}\tc) \ .
\label{l1}
\ee
$\lm_1$ is obtained by solving this equation.
$\lm_2$ is obtained by solving numerically $\Lamdar=0$ for the situation
$w(0)=0$.

\begin{figure}[hbt]
\centerline{\psfig{figure=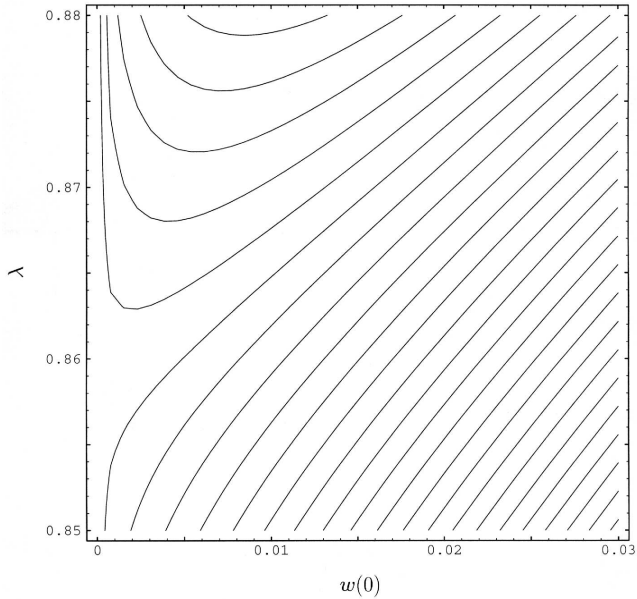,width=\figwidth}}
\caption{Contour plot of $k^2$ for $d=0.2$ and $\tc=\pi/2$ near P$_2$.
$k^2=\reva$ along the line $w(0)=0$. For $\lm \le \lm_2=0.858$, $k^2$ 
decreases with $w(0)$; for $\lm>\lm_2$, $k^2$ ``goes over a hill".}
\label{fig4}
\end{figure}

Fig.~\ref{fig3} shows $\lm_1$ and $\lm_2$ as functions of $d$ for
$\tc=\pi/2$. The
stability domain of the SC state increases monotonically with 
nonuniformity, following the trend given by the asymptotic formulae 
(\ref{define}). Fig.~\ref{fig4} shows a contour plot of $k^2$ as function
of $w(0)$
and $\lm$ for $\tc=\pi/2$ and $d=0.2$. This plot looks as if had been copied
from that obtained in Ref.\ \onlinecite{SIAM} for small
nonuniformity. For $\lm\le\lm_2$,
$k^2$ increases monotonically as $w(0)$ decreases; the value $|k|=\half$ is
reached when the order parameter describes the SC state, $w(0)=0$. 
On the other hand, for $\lm>\lm_2$, the value $|k|=\half$ is reached while
$w(0)$ is still positive; after that, for the minimum of the free energy,
$k$ will jump to $-k$ when the magnetic field is swept, so that lower values 
of $w(0)$ can only be achieved by metastable states.

\subsection{Dependence of $\lm_{1,2}$ on $\tc$}

\begin{figure}[hbt]
\centerline{\psfig{figure=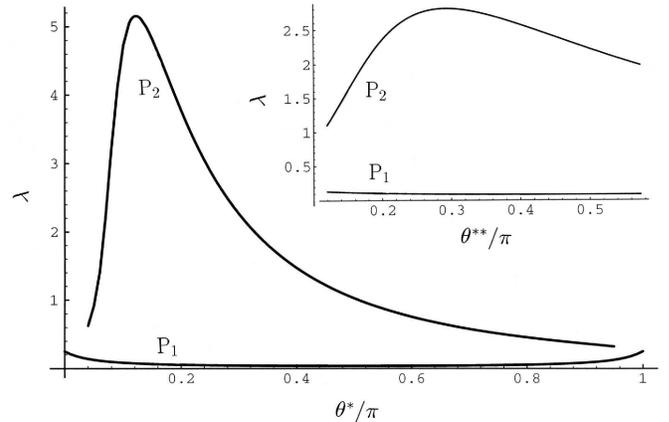,width=\figwidth}}
\caption{Position of the critical points as functions of the fraction of the
loop  where $D$ is small. Here $D(0)/D(\pi)=d=0.1$. The inset shows these
positions for the shape given by Eq.~(\protect{\ref{thsq}}), with $p=2$.
Again, $D(0)/D(\pi)\approx d=0.1$.}
\label{fig5}
\end{figure}

Fig.~\ref{fig5} shows $\lm_1$ and $\lm_2$ as functions of $\tc$ for
$d=0.1$. The curve
for $\lm_1$ has a maximum deviation from $\reva$ for $\tc=\pi/2$, and is
symmetric about this point. This is qualitatively what one would expect from
(\ref{define}). However, the curve for $\lm_2$ has a marked maximum for 
$\th<<\pi$. This suggests that large domains of stability for the 
SC state may be found for weak links, provided that the 
dimensions of the link are properly tuned. The limit $d,\tc<<1$ will be 
studied in Sec.\ III.

\subsection{Location of the Minimum of the Order Parameter}

According to Eq.~(\ref{zero}), which was derived in Sec.~6 of
 Ref.\ \onlinecite{SIAM}, the angle where $y(\th)$ has a minimum and
may eventualy 
vanish depends only on the geometry, and not on the temperature or the 
magnetic field. We want to check whether this result remains true when
 $D(\th)$ is far from uniform. The case described by Eq.~(\ref{piece}) is 
not appropriate for this purpose, since the minimum will always be at 
$\th=0$ by symmetry.

\begin{figure}[hbt]
\centerline{\psfig{figure=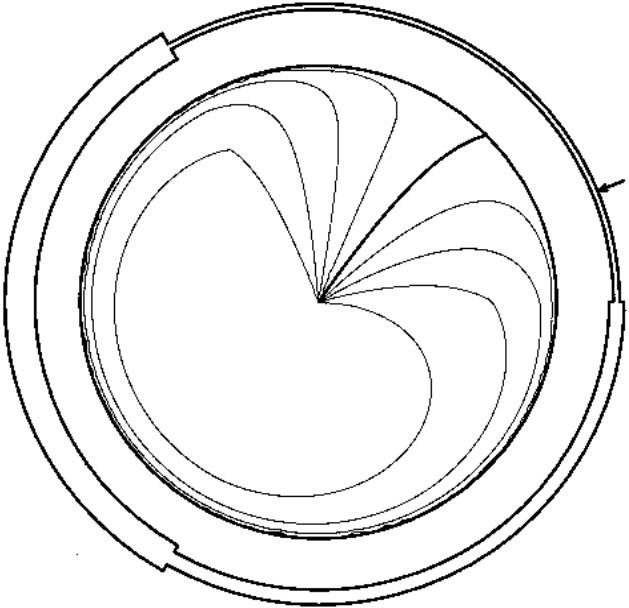,width=\figwidth}}
\caption{Outside: an asymmetric loop. Inside: contour plot of the order
parameter in this loop for $k^2=\reva$.}
\label{fig6}
\end{figure}

As a representative example, we consider a ring formed by three pieces of 
equal length, with cross sections in the ratio $2:10:5$, as shown in the 
outer part of Fig.~\ref{fig6}. We have calculated $y(\th;\lm)$ along the
line $\Gm_{\rm III}$ (see Fig.~\ref{fig2}), where $y$ vanishes for some
$\th$.

The inner part of Fig.~\ref{fig6} shows a polar contour plot of
$y(\th;\lm)$. The
radial coordinate is $\lm_2-\lm$, whereas the angular coordinate is just the 
angle $\th$ along the ring. For $\lm=\lm_1$, $y(\th)$ vanishes for all $\th$
(outer circumference); for $\lm>\lm_1$, $y(\th;\lm)=0$ along the nearly radial
line on the upper right. As $\lm$ increases, the layer where $y(\th)=0$
moves towards the middle of the thin piece of the ring. This could be
expected intuitively since, the smaller the coherence length, the less the
influence of the neighboring pieces. The arrow at the right shows the angle
where the order parameter would vanish according to the asymptotic result
(\ref{zero}). We see that the actual result is much less influenced by the
neighboring pieces; this tells us that regions where $D(\th)$ is very small
have a stronger influence than what is predicted if only their deviation from 
the average thickness is taken into account.

We have also evaluated the order parameter along the line $\Gm_{\rm I}$. As
we go from P$_1$ towards $\lm=k^2=0$, the minimum of the order parameter
moves towards the middle of the thin piece, but only by a very small angle.

\section{WEAK LINKS AND SIMILARITY}

We want to evaluate the asymptotic value of $\lm_2$ for $d,\tc<<1$, near the
region where $\lm_2$ is maximum. We shall also assume (and verify a 
posteriori that this is a self consistent assumption) that $\lm_2>>1$ and
$\tc/d$ remains finite in this regime. Our initial goal is the evaluation of
$\Lm_{\rm R}$ and, for it, we require $w(\th)$. $w(\th)$ is obtained from
(\ref{elw}), with $w(0)=1/\Lm=0$. Due to the continuity of $Dw'$,
$w'({\tc}^+)<<w'({\tc}^-)$. This boundary condition, together with $\lm>>1$,
make it reasonable to assume $w(\th) \approx 1$ in the interval 
$\tc\le\th\le\pi$. In the interval $0\le\th\le\tc$ we write 
$x=\sqrt{\lm}\th$,
$w(\th)=\bar w(x)=\bar w(\sqrt{\lm}\th)$, and (\ref{elw}) becomes
\be
\bar w \bar w''-\half\bar w'^2+2\bar w^2(1-\bar w)=0,
\label{bar}
\ee
where the derivatives are with respect to $x$ and the boundary conditions are
\be
\bar w(0)=0, \;\; \bar w(\bar\th)=1,
\label{barcond}
\ee
where $\bar\th=\sqrt{\lm}\tc$. The similarity solution of 
(\ref{bar})--(\ref{barcond}), $\bar w(x)$, is independent of $d$ and 
depends on $\lm$ and $\tc$ only through the combination $\bar\th$. 

By algebraic manipulation, we rewrite (\ref{Lamdar}) as
\be
\Lm_{\rm R}=2\int_0^{\tc}\frac{\tilde{f}-f}{f\tilde{f}}d\th-
2\int_{\tc}^\infty \frac{d\th}{\tilde{f}}+2\int_{\tc}^\pi\frac{d\th}{f} \ .
\label{Lambar}
\ee
Noting that $\tilde{f}(\th)=\half d\bar w''(0)x^2$, that $f(\th)=1$ for
$\th\ge\tc$, and changing the variable of integration, this becomes
\begin{eqnarray}
\Lm_{\rm R}&=&\frac{2}{\bar w''(0) d\sqrt{\lm}}\left(\int_0^{\bar\th}
\frac{\bar w''(0)x^2-2\bar w(x)}{\bar w(x)x^2}dx-\frac{2}{\bar\th} \right)
\nonumber \\
&&+2(\pi-\tc) \ .
\label{Lambar1}
\end{eqnarray}
At P$_2$, $\Lm_{\rm R}=0$. Using (\ref{Lambar1}) and $\tc<<\pi$ we obtain
\be
d\sqrt{\lm_2}=\frac{1}{\pi\bar w''(0)}\left(
\frac{2}{\bar\th}
-\int_0^{\bar\th}\frac{\bar w''(0)x^2-2\bar w(x)}{\bar w(x)x^2}dx \right) \ .
\label{universal}
\ee
Eq. (\ref{universal}) gives a universal expression for the scaled 
$\sqrt{\lm_2}$ as a function of $\bar\th$ only,  for any $d$ which is 
sufficiently small. Note that $\bar\th$ enters (\ref{universal}) not only
through the limits of integration, but also through (\ref{barcond}).

\begin{figure}[hbt]
\centerline{\psfig{figure=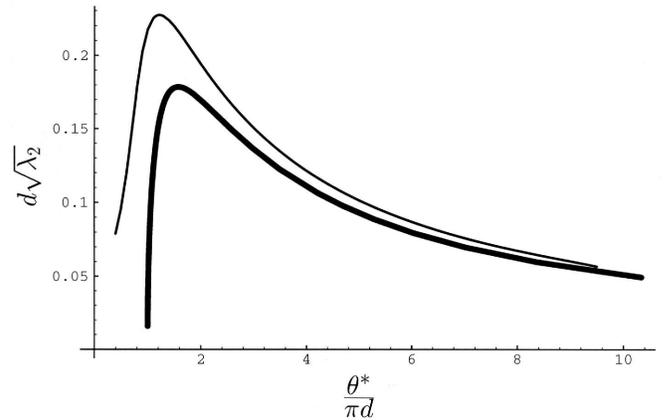,width=\figwidth}}
\caption{Position of P$_2$ for the shape (\protect{\ref{piece}}) with
$d<<1$. Thick 
line: asymptotic line, given by (\protect{\ref{universal}}); in this case 
$d\protect{\sqrt{\lm_2}}$ depends on $\tc/d$ only. Thin line: $d=0.1$.}
\label{fig7}
\end{figure}

The solution of (\ref{bar}) is of the form
\be
\bar w(x)=\frac{2m}{1+m}{\rm sn}^2(\frac{x}{\sqrt{1+m}},m) \ ,
\label{solbar}
\ee
where sn is a Jacobian elliptic function.
Given $\bar\th$, $m$ is obtained by numerical solution of (\ref{barcond}) and
the right hand side of (\ref{universal}) can be evaluated. After 
$d\sqrt{\lm_2}$ is known, we also obtain $\tc/d=\bar\th/(d\sqrt{\lm_2})$.
In this way we have obtained the universal curve for P$_2$ in
Fig.~\ref{fig7}, where
$\bar\th$ has been swept as a parameter. We see that $\lm_2$ is of the order
of $1/d^2$, provided that $\tc/d$ is not too small. If this ratio is below
$\sim\pi$, the assumption that $w$ almost reaches 1 at $\th=\tc$ is no longer
justified.

We are interested in large values of $\lm_2$. Numerical maximization of
(\ref{universal}) gives $d\sqrt{\lm_2}=0.179$ at $\bar\th=0.883$. Denoting
by $\tc_{\rm opt}$ the value of $\tc$ for which $\lm_2(d)$ is maximum, this
yields $\tc_{\rm opt}=1.57 \pi d$ for $d<<1$. On the other hand, 
(\ref{define}) gives $\tc_{\rm opt}=\pi/2$ for $d\approx 1$. A simple 
interpolation formula which has been found to give good results throughout is
\be
\frac{\tc_{\rm opt}}{\pi}=\frac{0.734d}{0.468+d} \ .
\label{opt}
\ee

We shall close this section by considering the behavior of the loop for
``typical experimental conditions" near the SC state. For given $d$, we would
like $\lm_2-\lm_1$ to be as large as possible; for this purpose, a reasonable
choice is $\tc=\tc_{\rm opt}$. A typical temperature would be in the middle
of the stability domain of the SC state. We therefore define
$\lm_{\rm typ}(d)=(\lm_1(\tc_{\rm opt})+\lm_2(\tc_{\rm opt}))/2$.
Let us now fix $\tc=\tc_{\rm opt}$ and $\lm=\lm_{\rm typ}$ and sweep the
magnetic flux (which determines $k$). As a response, the supercurrent 
$I=k/\Lm$ varies. It vanishes for $k=0$ and for $k=\half$, and reaches its 
maximum $I=I_{\rm max}$ for
some intermediate value $k=k_{\rm max}$. In order to characterize the 
behavior of the loop, we have evaluated $k_{\rm max}(d)$. 

\begin{figure}[hbt]
\centerline{\psfig{figure=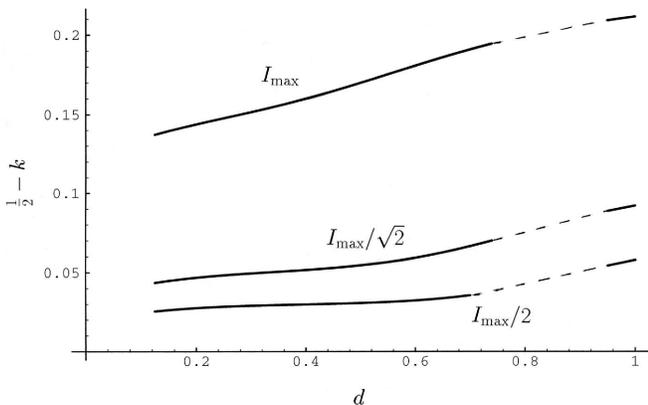,width=\figwidth}}
\caption{Rate at which the supercurrent decreases when the SC state is
approached
under ``typical experimental conditions". Calculations have been performed
in the region where $m_1$ in (\protect{\ref{l*}}) is complex. For $d\ge
0.95$ we have used the asymptotic results developed in
Ref.\ \protect{\onlinecite{SIAM}}.}
\label{fig8}
\end{figure}

The criterion to determine $k_{\rm max}(d)$ is as follows. Consider the 
system of equations that determines $\nu_i$ and $m_i$ in Subsec.~II.A.
Note that $k$ enters these equations only through the constant term in 
(\ref{elcons}) and that this term is an increasing function of $I$. 
Therefore, this constant term has a maximum possible value, 
which is attained at $k_{\rm max}$. Above this value
the system of equations has no solutions and, below it, it has two (one for
$k<k_{\rm max}$ and one for $k>k_{\rm max}$). At $k=k_{\rm max}$ the system
of equations has a double root and, therefore, its Jacobian vanishes. This
provides the additional equation from which $k_{\rm max}(d)$ can be 
determined. The results are shown in Fig.~\ref{fig8}. Also shown are the
values of
$k$ at which $I$ has decreased by the factors $\sqrt{2}$ and 2, as the
SC state is approached. These are obtained by fixing the constant term in
(\ref{elcons}).

Fig.~\ref{fig8} shows that the supercurrent decreases sharply as the SC
state is approached. This effect might be used for calibration.

\section{DEVIATIONS FROM PIECEWISE CONSTANT THICKNESS}

Sections II and III answer our questions for the special case that $D(\th)$
is of the form (\ref{piece}). We ask now whether the behavior that has been
found is typical of loops that resemble a weak link, or arises from spurious
features of the model, such as the discontinuity at $\th=\tc$. For this 
purpose, we consider $D$ of the form
\be
D(\th)=d+(1-d)\tanh^p(\th/\Tc)
\label{thsq}
\ee
for $0\le\th\le\pi$ and symmetric about 0 and $\pi$. For $\Tc<<\pi$ this 
shape models a weak link with $D(0)/D(\pi)=d$. The larger the value of $p$,
the more the shape of $D$ will resemble a piecewise constant form.

\subsection{Smooth Thickness at $\th=0$} 

We have investigated the case $p=2$ in Eq.~(\ref{thsq}).
This time the differential equations in (\ref{ely}) or (\ref{elw}) have to
be solved numerically. $\lm_1$ is found by linearizing (\ref{ely}), assigning
an arbitrary normalization to $y'(0)$, and looking for the smallest value of 
$\lm$ that
gives a periodic solution. $\lm_2$ is found by requiring $\Lm_{\rm R}=0$.

In order to evaluate $\Lm_{\rm R}$, we have to compute $y(\th)$. For this
we solve (\ref{ely}) with the boundary conditions $y(0)=y'(\pi)=0$. This is
still unsufficient, since we need the ratio of the strongly vanishing terms
$\tilde f- f$ and $\tilde f f$ near $\th=0$. For this purpose, a power series
for $y(\th)$ was developed near $\th=0$.

\begin{figure}[hbt]
\centerline{\psfig{figure=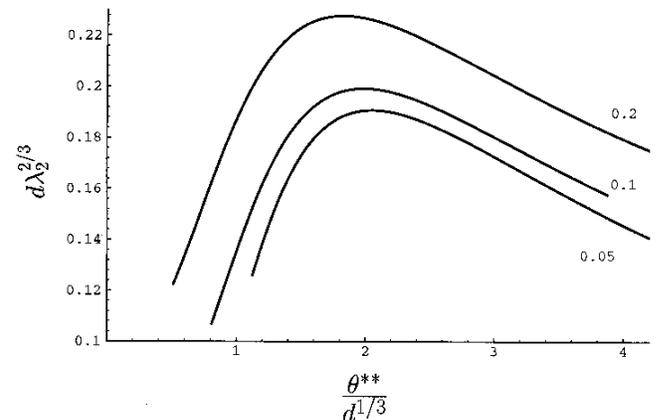,width=\figwidth}}
\caption{Position of P$_2$ for the shape (\protect{\ref{thsq}}), with
$p=2$. The lines
are marked by the values of $d$. The scaling has been chosen empirically to
mimic a nearly universal behavior, as in Fig.~7.}
\label{fig9}
\end{figure}

The inset in Fig.~\ref{fig5} shows $\lm_{1,2}$ as functions of $\Tc$ for
$d=0.1$. Fig.~\ref{fig9} is an empirically scaled set of graphs for
$\lm_2(\Tc)$. The 
similarity obtained in Sec.~III is no longer obeyed, but the qualitative
results of the piecewise constant case are still true: for fixed $d$, 
$\lm_2(\Tc)$ has a maximum for some shape of the thickness,
$\Tc=\Tc_{\rm opt}$; as $d$ decreases, the peak of the unscaled $\lm_2(\Tc)$ 
becomes sharper, with larger $\lm_2(\Tc_{\rm opt})$
and smaller $\Tc_{\rm opt}$; for $\Tc<\Tc_{\rm opt}$, $\lm_2$ drops sharply.

\subsection{$D'(0) \neq 0$}

We have investigated the case $p=1$ in Eq.~(\ref{thsq}), which deviates
stronger from piecewise constant than the case $p=2$. The numerical analysis
is very similar to that of the previous section, with one additional 
subtlety: strictly, the quantity $\Lm_{\rm R}$ defined by (\ref{f}) and
(\ref{Lamdar}) should not be evaluated with the function $w(\th)$ which
describes the SC state. Rather, we should evaluate it with $w(0)>0$ and
take the limit $w(0) \ra 0$. Let us define by $\Lm_{\rm R,SC}$ the value
of $\Lm_{\rm R}$ which is obtained by using the singly-connected order
parameter in Eq.~(\ref{f}). Separating $\Lm_{\rm R}$ into a part which is
continuous when going to $w(0) \ra 0$ and a remainder, we obtain
\be
\lim_{w(0) \ra 0}\Lm_{\rm R}=\Lm_{\rm R,SC}-\frac{D'(0^+)-D'(0^-)}
     {D(0)^2w''(0)} \ .
\label{limit}
\ee
If $D$ is smooth at $\th=0$, the last term vanishes.

\begin{figure}[hbt]
\centerline{\psfig{figure=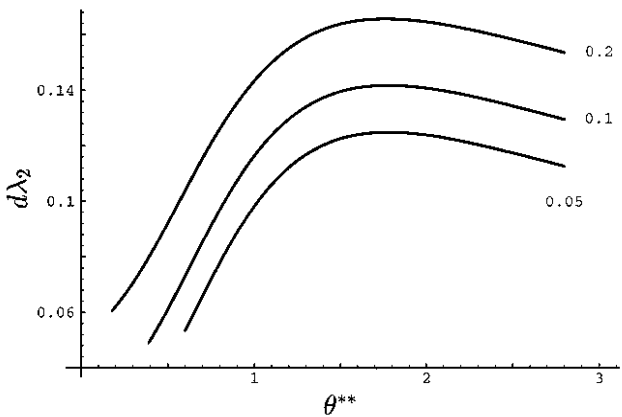,width=\figwidth}}
\caption{Like Fig.~9, for $p=1$.}
\label{fig10}
\end{figure}

Using this formula, we have calculated the curves in Fig.~\ref{fig10}. As
might 
already be expected, these curves have maxima for suitable $\Tc$, and $\lm_2$
increases when $d$ decreases, but the peaks are less pronounced and the
dependence on $d$ is weaker than in the previous examples.

\section{DISCUSSION}

The examples considered here extend the results of
 Ref.\ \onlinecite{SIAM} to cases
where the thickness is far from uniform. The critical points P$_1$ and P$_2$
still exist and behave as in Ref.\ \onlinecite{SIAM}. The stability domain
of the
singly-connected state increases monotonically with nonuniformity. The 
dependence of $\lm_1$ on $\tc$ is symmetric about $\pi/2$ and in qualitative
agreement with Eq.~(\ref{define}).

There is a surprise in the dependence of $\lm_2$ on $\tc$. 
For the thickness profile defined in (\ref{piece}), $\lm_2(\th)$ has a
sharp
peak when the ring has a ``weak link" ($d,\tc<<1$) and the ratio $\tc/d$ is
appropriately tuned, as found in Sec.~III. In this case, the length of the
temperature domain where the SC state is stable is inversely proportional to 
$d^2$.

Modern versions of the Little-Parks experiment are performed on mesoscopic
loops, produced by microlitographic techniques. Smaller values of the 
perimeter $2\pi R$ allow for smaller values of the coherence length and
thus expand the relevant temperature range below $T_c$. The basic mesoscopic
loop is usually repeated a large number of times
(e.g. Ref.\ \onlinecite{exp}), but there are also cases in which a single
loop is 
used (e.g. Ref.\ \onlinecite{moch2}). The main contribution
of this paper is the design of the thickness profile of the basic mesoscopic
ring.

Our goal is to design a sample with a large temperature domain for the SC 
state. Among the situations considered, we have seen that
this is best achieved with the piecewise constant profile (\ref{piece}),
with $\tc=\tc_{\rm opt}$. If $\tc_{\rm opt}$ is not accurately known, it is
safer to take $\tc>\tc_{\rm opt}$ than $\tc<\tc_{\rm opt}$. The temperature
domain can be increased by decreasing the ratio $d=D(0)/D(\pi)$.

The piecewise constant thickness poses a mathematical problem. Our one
dimensional free energy in Eq.~(\ref{gl3}) is based on the assumption
that the order parameter $\psi$ and the supercurrent density 
vector depend only on $\th$ and remain constant on
every given cross section. If $D(\th)$ is smooth, it has been proven that
this assumption is approached when the thickness (linear dimension of $D$)
is much smaller than the perimeter of the loop. But a piecewise constant 
thickness is not smooth; near the region where the thickness changes, the
streamlines are strongly curved and the assumption that the current density
remains constant on the cross section is an unrealistic picture.
Nevertheless, the results of Sec.~IV add confidence to the results of the 
piecewise constant case as a plausible limit.
In order to have a safer answer to the piecewise constant case, and also
since the usual experimental situation is that the thickness is smaller than
the perimeter only by a moderate ratio, a two dimensional treatment is 
desirable. Even if the results for the piecewise constant case turned out to
stem from an over simplified free energy functional, the results of Sec.~IV 
indicate that the temperature domain where the SC state is stable increases
at least as the ratio $D(\pi)/D(0)$. 

It is experimentally possible to make superconductig samples with weak links, 
such that $D(0)$ is smaller than $D(\pi)$ by several orders of magnitude. 
However, in this case our theoretical treatment becomes inappropriate. 
The most important reason is that a large ratio $D(\pi)/D(0)$ implies a 
large volume where the streamlines are strongly curved and a large 
influence of the limitation discussed in the previous paragraph.
 Second, if the coherence length is much smaller than the perimeter, it may 
not necessarily be much larger than the thickness. And third, far 
from $T_c$ the Ginzburg-Landau theory is not necessarily a good 
approximation. Moreover, it is not necessary to have $d<<1$; a moderately
weak link such as $d \sim 0.1$ will already give a temperature domain 
which is about twenty times larger than that of the ``classic" Little-Parks
effect ($0 \le \lm \le \reva$), and should be easily accessible. 

Expression (\ref{gl3}) for the free energy neglects the contribution of the
magnetic field due to the supercurrent. This contribution was evaluated in
Ref.\ \onlinecite{SIAM} and it enhances the temperature
domain for the SC
state. This contribution may be important for a large periodic array of rings
at $\lm>\lm_2$ and $|k|=\half$;
 in this case the ground state due to (\ref{gl3}) is 
degenerate and a small perturbation may give rise to collective effects.

Several possibilities for identification of the SC state and the critical 
point P$_2$ have been discussed previously.\cite{beru,SIAM} One
possibility is to 
observe directly whether the order parameter vanishes at some layer $\th=0$, 
by means of STM or some decoration technique. Another possibility is to check
whether the supercurrent vanishes or jumps for $|k| \approx \half$; this can
be done by means of magnetometer, as in Ref.\ \onlinecite{silver}. The SC
state was not
observed in Ref.\ \onlinecite{silver}; conceivably, the effective value of
$\tc/d$ was
too low.

For $\lm>\lm_2$ and $|k|=\half$, the free energy is a bistable potential.
The potential barrier decreases and the overlap between the order parameters 
at the minima increases as $\lm$ approaches $\lm_2$. Therefore, fluctuations
are expected between both minima, which occur more and more frequently as
$\lm \ra \lm_2^+$. However, for $\lm\le\lm_2$ the free energy as only one
minimum, and no fluctuations are expected. This effect was indeed found in
Ref.\ \onlinecite{osc}, but it was understood that the sample is in the
normal state 
after the fluctuations disappear. According to the present interpretation,
at this stage the sample is in the SC state and is still superconducting.
We have seen that P$_2$ can be pushed to a regime where superconductivity 
is well developed, so that this bistable potential might open the 
possibility for low dissipation logical devices.

If P$_2$ is pushed to a sufficiently low temperature and thermal 
excitations are rendered unimportant, one may expect to 
achieve macroscopic quantum coherence \cite{leg} of both minima in the 
bistable potential. This situation was looked for in the past with 
Josephson junctions.\cite{coh} The setup described in this article seems 
to be cleaner and more promising: there are no uncontrollable parts in the
loop, no external current and less sources of dissipation. 

To summarize: our results indicate that the temperature region where the 
singly-connected state exists is not necessarily small, and can be readily 
found with existent experimental techniques.

\begin{center} {\bf ACKNOWLEDGMENTS}  \end{center}

This research was supported by the US-Israel Binational Science Foundation.
J.\ B.\ was also supported by the Israel Science Foundation.

\end{document}